\documentclass[nofootinbib, twocolumn,showkeys, amssymb]{revtex4-1}
\usepackage{amsfonts,amsmath,amssymb,amsthm,hyperref}
 
\usepackage{color,psfrag}
\usepackage[dvips]{graphicx}

\newcommand{\be}{\begin{equation}}
\newcommand{\ee}{\end{equation}}
\newcommand{\ba}{\begin{eqnarray}}
\newcommand{\ea}{\end{eqnarray}}

\begin{document} 

\title{Towards Loop Quantum Supergravity (LQSG)}

\author{N. Bodendorfer$^1$}
\email[]{norbert.bodendorfer@gravity.fau.de} 

\author{T. Thiemann$^{1}$}
\email[]{thomas.thiemann@gravity.fau.de}

\author{A. Thurn$^1$}
\email[]{andreas.thurn@gravity.fau.de}

\affiliation{$^1$Inst. for Theoretical Physics III, FAU Erlangen -- N\"urnberg, Staudtstr. 7, 91058 Erlangen, Germany}

\begin{abstract}
Should nature be supersymmetric, then it will be described 
by Quantum Supergravity at least in some energy regimes. The currently 
most advanced description of Quantum Supergravity and beyond is Superstring
Theory/M-Theory in 10/11 dimensions. String Theory is a top-to-bottom 
approach to Quantum Supergravity in that it postulates a new object, the 
string, from which classical Supergravity emerges as a low energy limit. On the other hand, one may try more traditional bottom-to-top routes and 
apply the techniques of Quantum Field Theory. Loop Quantum Gravity (LQG) is a manifestly 
background independent and non-perturbative approach to the quantisation
of classical General Relativity, however, so far mostly without 
supersymmetry. The
main obstacle to the extension of the techniques of LQG to the quantisation 
of higher dimensional Supergravity is that LQG rests on a specific 
connection formulation of General Relativity which exists only in D+1 = 4
dimensions. In this Letter we introduce a new connection formulation of General Relativity
which exists in all space-time dimensions. We show that all LQG techniques 
developed in D+1 = 4 can be transferred to the new variables in all dimensions
and describe 
how they can be generalised to the new types of fields that appear in 
Supergravity theories as compared to standard matter, specifically 
Rarita--Schwinger and p-form gauge fields.
\end{abstract}


\maketitle

String/M Theory (ST) \cite{GSW,Polchinski} and Loop Quantum Gravity (LQG)
\cite{CarloBook,TTBook} are rather different programmes that aim 
for a consistent synthesis of the principles of General Relativity and 
Quantum Theory. String/M Theory is necessarily 10/11-dimensional, necessarily
supersymmetric and is perturbatively defined on appropriate 
background space-times. Loop Quantum Gravity, on the other hand, is to date 
restricted to 4 space-time dimensions, does not need supersymmetry and 
by design is background independently and non-perturbatively defined.
It is therefore very hard to compare these two approaches. 

One possibility to make contact between them 
consists in the consideration of String/M Theory on space-time manifolds for 
which the excess dimensions are compactified and in regimes where 
supersymmetry is broken, so that only an effective 4-dimensional theory of 
General Relativity and the Standard Model plus quantum corrections 
survives, which then can be compared to the sector of LQG with small 
geometry fluctuations around the chosen background. We refer to 
\cite{STPheno} for the state of the art of String Theory Phenomenology
but it transpires that there are many possibilities for doing 
this and there appears to be no specific model that one can compare LQG
to. 

Another possibility
would be to generalise LQG to higher dimensions including supersymmetric 
matter in order to compare the methods of the two theories directly in 
10/11 dimensions. This idea appears to be easier to implement 
because in its fundamental 
dimension, String/M Theory is much simpler to describe and there are 
much less choices to be made. Concretely, 
String Theory can be considered as a specific proposal for quantising 
classical Supergravity \cite{SUGRA} in 10/11 dimensions on a background defined by 
a solution to the classical Supergravity equations of motion.
With the possible exception of $N=8$
Supergravity in 4 space-time dimensions \cite{D=4N=8,D=4N=8a}, perturbative approaches seem 
to fail due to non-renormalisability\footnote{More precisely, to date
no counterterm has been found for $N=8$ SUGRA in 
4 dimensions in any loop contribution carried out so far, some of them 
way beyond 3 loop order. This could mean that the theory is perturbatively
finite like QED in 4 dimensions. It does not mean that it is a UV completion
of General Relativity because the perturbation series most 
likely diverges. Therefore strictly speaking, also here non-perturbative 
methods need to be employed.} whence non-perturbative methods 
appear to be more promising. 
Thus, the natural question arises, how to apply non-perturbative methods, such
as those of LQG,
to the quantisation of classical Supergravity in 10/11 dimensions. This is a subject
of obvious interest to Supergravity research, 
but, to the best of our knowledge, the literature on it is rather sparse 
\cite{NicolaiMelosch}.

We caution the reader that although new steps towards a loop quantisation of higher dimensional Supergravity will be taken in this Letter, there are several issues left to deal with. One of them is to gain a proper understanding of the quantum dynamics. Although well-defined operators corresponding to the classical Hamiltonian and supersymmetry constraints can be constructed, there is at the moment insufficient control over their algebra, in particular off shell. A faithful representation of the super Dirac algebra on the other hand might be considered as one of the benchmarks a quantisation of a Supergravity has to satisfy. Also, it is presently unclear how one could establish a relation to string theory which goes beyond comparing two quantisation schemes for Supergravity. We hope that first hints can be obtained by looking at situations which can be treated by both theories non-perturbatively, like the calculation of black hole entropy. 

The results of this Letter can be summarised as follows: We derive a generalisation of Ashtekar variables in all dimensions $D+1 \geq 3$ with the properties necessary for a rigorous quantisation. Furthermore, we generalise the construction to standard model matter, Rarita-Schwinger fields\footnote{In the present treatment, it is necessary to have a real representations of the Lorentzian Clifford algebra. We thus have to restrict to such dimensions, however, the most interesting cases $D+1=4,10,11$ are covered.}, and Abelian p-forms, and thus to the matter fields appearing in many interesting Supergravities. A Hilbert space representation for a suitable point splitting Poisson subalgebra (the holonomy--flux algebra extended to the above matter fields) is provided and densely defined operators corresponding to the constraints of the classical theory can be constructed. The issue of a non-anomalous representation of the constraint algebra has not been investigated so far, however, these problems can, in principle, be circumvented using a Master constraint treatment \cite{Master}. Still, having a faithful representation of the classical constraint algebra would be far more satisfactory and we hope to revisit this issue in the future.\\
\\
When trying to find a generalisation of Ashtekar's variables to higher dimensions, one almost immediately gets stuck:\\
LQG in 4 dimensions is fundamentally based on a connection formulation 
of the gravitational degrees of freedom. However, the construction
of this connection and the properties it has and which make it so 
adapted for purposes of quantisation, exist only in 4 space-time 
dimensions \cite{Ashtekar, Barbero, Immirzi}. Specifically, in 4 dimensions,
General Relativity can be described, in its Hamiltonian form, by an SU(2)
Yang--Mills phase space, subject to Gau{\ss}, spatial diffeomorphism 
and Hamiltonian constraints. The key properties of this formulation are:
\begin{itemize}
\item[(I)] The connection $A_a^j$ and its canonically conjugate momentum $E^a_j$
(Yang--Mills ``electric field'') are real-valued, the $^\ast$-relations 
of the associated Poisson algebra are trivial. Here $a,b,c,\hdots=1,2,3$ 
are spatial, tensorial indices and $j,k,l,\hdots=1,2,3$ are su(2) indices. 
\item[(II)] The Poisson algebra is of the usual, simple CCR form
\ba 
\{A_a^j(x),A_b^k(y)\}=\{E^a_j(x),E^b_k(y)\}=0 \nonumber \\
\{E^a_j(x),A_b^k(y)\}=-G\gamma\delta^a_b\delta_j^k \delta(x,y) \text{,} \label{1}
\ea
where $G$ is Newton's constant and $\gamma$ is a free, real-valued parameter, called the Immirzi parameter \cite{Barbero, Immirzi}.
\item[(III)] The gauge group SU(2) is compact.
\end{itemize}
These three properties are essential for the whole LQG framework. Without
them, LQG would not exist. Properties (I) and (II) imply that there is 
a sufficiently simple $^\ast$-algebra $\mathfrak{A}$ of functions 
on phase space separating its points so that one has a chance to find non-trivial Hilbert space
representations thereof. Property (III) implies that the holonomies of 
$A$ are valued in a compact set. It is therefore possible to construct
a probability measure on the space of (distributional) connections.
The Hilbert space is then an $L_2$ space of functions of connections 
which makes sense due to (\ref{1}) because on such a space the 
connection acts by multiplication which is only consistent with the 
algebra if the connection is Poisson self-commuting.

Put together, this enabled to develop a rigorous kinematical 
mathematical framework \cite{AI,AL} and to find a background independent 
representation of the holonomy--flux $^\ast$-algebra $\mathfrak{A}$
which was later shown to be the unique one \cite{LOST,Fleischhack}
when one insists on a unitary representation of the spatial diffeomorphism
group. Moreover, this representation is also well adapted to the quantum 
dynamics as one may rigorously implement the spatial diffeomorphism constraint 
\cite{ALMMT} and the Hamiltonian constraint \cite{QSD1} including standard
matter \cite{QSD5,KinematicalHS} without anomalies\footnote{As in the $3+1$-dimensional treatment \cite{TTBook}, anomalies are absent in the following sense: The commutator of two Hamiltonian constraint operators acting on a spin network state is annihilated by diffeomorphism invariant distributions, which mimics the classical Dirac algebra relation $\{C, C \} = \vec{C}$. In which sense an off-shell closure is possible is presently unknown, since due to the non-continuous representation used, an operator corresponding to infinitesimal spatial diffeomorphism does not exist, and only a representation of finite diffeomorphisms can be defined in the quantum theory.}.   
 
It transpires that LQG heavily relies on a connection formulation 
with the properties listed above. This connection can be found by two 
independent methods. The first stays purely within the Hamiltonian 
framework \cite{Ashtekar}
and uses an extension of the ADM phase space of General Relativity 
\cite{Wald} to the afore mentioned Yang--Mills phase space which is subjected
to the additional su(2) Gau{\ss} constraint in order that its symplectic
reduction recovers the ADM phase space. The core of the proof of this 
so-called {\it symplectic reduction theorem} is the observation, that the 
spin connection of the triad $e^a_j=E^a_j/\sqrt{|\det(E)|}$ has a potential, 
that is, there exists a functional $F[E]$ such that 
$\Gamma_a^j(x)=\delta F/\delta E^a_j(x)$. The second method uses 
the Lagrangian framework and starts from the Holst generalisation
\cite{Holst,BarroeSa} of the Palatini action in order to accommodate the 
Immirzi parameter. This is actually an SO(1,3) Yang--Mills theory
phase space in Lorentzian signature, however,
it is subject to second class constraints which in particular imply that
the connection is not self-commuting with respect to the corresponding
Dirac bracket \cite{Alexandrov}. Therefore, properties II. and III. 
listed above are not satisfied and no kinematical Hilbert space representation
of the Dirac bracket algebra has been found so far. In order to obtain the 
SU(2) Yang--Mills phase space without second class constraints, one therefore
imposes the so-called time gauge which fixes the boosts of SO(1,3) and solves
the second class constraints. What remains from the so(1,3) connection
is the afore mentioned su(2) connection whose Dirac brackets are the 
Poisson brackets (\ref{1}).

One would now guess that one can simply repeat either of these methods 
in dimensions $D+1>4$. However, this is not the case. As has been shown 
in \cite{Han}, the second route leads to a SO$(D)$ gauge theory in the 
time gauge but it is not
a theory with a connection. In order to obtain a connection formulation one 
would therefore need the analog of a topological Holst term, but such a term
is not available in all dimensions. Without time gauge, one does have an SO$(1,D)$ 
connection formulation but subject to a Dirac bracket which leads to the 
same complication as in $D+1=4$. Moreover, the gauge group SO$(1,D)$ is not
compact so that the functional analytic and measure theoretic tools mentioned
before do not apply. The first route also meets difficulties:\\
In $D$ spatial dimensions, the metric of the ADM phase space has $D(D+1)/2$
degrees of freedom while an so$(D)$ connection has $D^2(D-1)/2$ degrees 
of freedom of which the so$(D)$ Gau{\ss} constraint fixes ${D(D-1)/2}$. The 
symplectic reduction of the Yang--Mills phase space by the Gau{\ss} constraint
therefore leaves ${D^2(D-1)/2-D(D-1)/2=D(D-1)^2/2}$ degrees of freedom
which equals $D(D+1)/2$ precisely for $D=3$. If one wants to go beyond 
$D=3$ one therefore must add more constraints and/or change the 
gauge group in order to match 
the correct amount of ADM degrees of freedom and these constraints should 
better be first class in order to avoid complicated Dirac brackets and 
associated non-commuting connections.

The general analysis of the first route has been started in 
\cite{Peldan,Peldan2}. We take an unbiased viewpoint and consider a 
general Yang--Mills phase space with gauge group $G$, connection  
$A_a^\alpha$ and conjugate momentum $\pi^a_\alpha$ where \mbox{$\alpha=1,\hdots,N=\dim(G)$}
denotes the Lie algebra index. There are $DN$ degrees of freedom in 
the connection of which the Gau{\ss} constraint removes $N$. If there 
are no other constraints then we must have ${(D-1)N=D(D+1)/2}$. The 
only positive integer solutions to this equation are 
$(D,N)=(2,3),(3,3)$ which again 
corresponds to 3-dimensional or 4-dimensional gravity respectively with gauge groups 
SO(1,2) and SO(3) respectively (with Lorentzian signature). Thus, necessarily
more constraints are required.  

As has been demonstrated in \cite{Peldan}, one can obtain GUT theories 
by varying $G$. We are for the time being only interested in General 
Relativity and ask for the group $G$ of minimal dimension that accomplishes 
all our requirements. To constrain the possible choices we try to 
follow as closely as possible the treatment of \cite{Ashtekar,Peldan} in $D=3$
and consider a ``square root'' $e_a^I$ of the spatial metric 
$q_{ab}=\eta_{IJ} e^I_a e^J_b$ where $I,J,K,\hdots=1,\hdots,n\ge D$. Here $\eta$
defines a $G$ invariant metric of signature $(p,q)$ with $q\ge D$ which is necessary for
$q_{ab}$ to be positive definite. This constrains the gauge group to
be SO$(p,q),\;\;n=p+q$ which has dimension $N=n(n-1)/2$. The basic idea of the proof of equivalence with the ADM formulation of General Relativity will be to define a gauge theory subject to certain constraints, the degrees of freedom of which will be related to the ADM degrees of freedom. One then needs to show that the symplectic reduction of the gauge theory coincides with the ADM phase space, i.e. the Poisson brackets of the ADM degrees of freedom coincide with the Poisson brackets derived form the gauge theory. 

Next to the spatial metric, we require an expression for the ADM momentum $P^{ab}:=\sqrt{\det(q)}q^{ac}[K_c\;^b-\delta_c^b \; K_d\;^d]$, where $K_{ab}$ is the extrinsic curvature. For this, we need to construct from $e_a^I$ the so-called 
hybrid connection $\Gamma_{aIJ}$ defined by 
\be \label{2}
D_a e_b^I=\partial_a e_b^I-\Gamma_{ab}^c e_c^I+\Gamma_a\;^I\mbox{}_J e_b^J=0
\ee
and, in close analogy to the $(3+1)$-dimensional case, we set
\be \label{3}
\sqrt{\det(q)}\;K_a\;^b:=[A_{aIJ}-\Gamma_{aIJ}]\pi^{bIJ} \text{,}
\ee
where $A$ is the so$(p,q)$ connection and $\pi$ its conjugate momentum. The logic behind this definition becomes clear only after the symplectic reduction to the ADM phase space has been performed: $K_{ab}$ will be shown to coincide with the usual extrinsic curvature from the ADM formalism on the constraint surface and the definition is thus consistent with the usual one. 
Notice that (\ref{3}) is meaningful since $A-\Gamma$ transforms as a
Lie algebra valued one form and not as a connection under SO$(p,q)$.
The question is of course whether $\Gamma$ exists. The fact that 
$\Gamma_{a(IJ)}=0$ leads to the consistency condition (all internal indices 
are moved with $\eta$)
\be \label{4}
e_{(c|I}\partial_{a|} e_{b)}^I-\Gamma_{(c|a|b)}=0 \text{,}
\ee
which can be shown to be identically satisfied. Therefore of the 
$D^2 n$ equations (\ref{2}) for the $Dn(n-1)/2$ coefficients $\Gamma_{aIJ}$ 
only $D^2 n-D^2(D+1)/2$ are independent. Requiring that $\Gamma_{aIJ}$ can
be uniquely solved for leads to a quadratic equation with the 2 roots
$n=D,\;n=D+1$, that is, either $p=0,q=D$ for $n=D$, or $p= 1, q=D$ and $p=0, q=D+1$ for $n=D+1$. 

Finally, the number of constraints additional to the Gau{\ss} constraint
needed is given by ${Dn(n-1)/2-n(n-1)/2-D(D+1)/2}$ which equals $D^2[D-3]/2$
for $n=D$ and $D(D+1)[D-2]/2$ for $n=D+1$. The question is of course what 
these constraints should be. A natural choice is that these constraints
should somehow impose that $\pi^{aIJ}$ is entirely determined by $e_a^I$. 
The excess number of degrees of freedom in $\pi^{aIJ}$ as compared to 
$e_a^I$ is given by $Dn(n-1)/2-Dn$ which equals precisely the number 
of additional constraints needed for both $n=D$ and $n=D+1$. In order 
to write $\pi^{aIJ}$ purely in terms of $e_a^I$ we require an internal 
vector $n^I$ built from $e_a^I$ such that 
\be \label{5}
\pi^{aIJ}=2\sqrt{\det(q)} q^{ab} n^{[I} e_b^{J]}   \text{.}
\ee
There is no way to construct $n^I$ out of $e_a^I$ algebraically for 
$n=D$ so that in this case we must resort to $D=3$. For $n=D+1$, however,
we may build the unit normal
\be \label{6}
n_I:=\frac{1}{D!\;\sqrt{\det(q)}}\; 
\epsilon_{IJ_1..J_D}\;\epsilon^{a_1..a_D}\; e_{a_1}^{J_1}.. 
e_{a_D}^{J_D}
\ee
satisfying $n_I n^I=- 1$ for SO$(1, D)$ or  $n_I n^I= 1$ for SO$(D+1)$.

We conclude that if we want to obtain a connection representation with 
compact gauge group for {\it Lorentzian General Relativity}, a natural
choice is to consider the phase space of an SO$(D+1)$ Yang--Mills theory subject
to an so$(D+1)$ Gau{\ss} constraint and additional {\it simplicity constraints}
which impose that $\pi^{aIJ}$ is determined by a generalised $D$-bein 
$e_a^I$ via (\ref{5}) where $n_I e_a^I=n_I n^I-1=0$ is the unit normal 
determined by $e_a^I$. In fact, one might have almost guessed that:\\
If one performs the Hamiltonian analysis of the Lorentzian Palatini action in 
$D+1$ dimensions (see e.g. \cite{II} and references therein) 
then one obtains a primary constraint precisely of the 
form (\ref{5}). However, as secondary constraints one obtains an so$(1,D)$ 
Gau{\ss} constraint
next to spatial diffeomorphism and Hamiltonian constraints plus one 
more constraint. This last constraint, let us call it $D$-constraint,
 is a second class constraint 
partner to the simplicity constraint $S$. Thus, the Hamiltonian analysis 
of the $D+1$ Palatini action for {\it Lorentzian gravity} 
fails to deliver a canonical theory of connections 
with the properties (I), (II) and (III) listed above for two reasons:\\
1. The gauge group is SO$(1,D)$ rather than SO$(D+1)$ and thus non-compact.\\
2. The theory suffers from second class constraints and thus leads 
to Dirac bracket non-commuting connections.

As it turns out \cite{II}, the second problem can be circumvented by 
employing the machinery of {\it gauge unfixing} \cite{GU1,GU2}. 
Under certain conditions, which are satisfied for our second class pair
$(S,D)$, it is possible to trade the second class system under consideration
for an equivalent first class system equipped with the original Poisson
bracket rather than the Dirac bracket so that the connection remains
Poisson commuting. However, the first problem cannot be overcome 
starting from the Palatini action for Lorentzian General Relativity.
Therefore, the canonical theory that we are about to describe does not
have an obvious Lagrangian origin (other than by backwards Legendre transform).

The first step \cite{I} consists in writing both the hybrid connection and 
the simplicity constraint purely in terms of $\pi^{aIJ}$. The 
guideline for doing this consists in replacing the solution 
$\Gamma_{aIJ}[e]$ of (\ref{2}), which can be computed explicitly,
by a function $\Gamma_{aIJ}[\pi]$ of $\pi$ alone 
such that it reduces to $\Gamma_{aIJ}[e]$ when $\pi^{aIJ}=2n^{[I} E^{a|J]} $
where $E^{aI}=\sqrt{\det(q)} q^{ab} e_b^I$. This has been done explicitly
in \cite{I} which leads to a rather complicated expression which 
can be displayed as a rational homogeneous function of degree zero 
in terms of $\pi$ and its first partial derivatives and which transforms 
as an so$(D+1)$ connection.

Next, one shows \cite{I} that the condition that $\pi^{aIJ}$ is of the 
form $2n^{[I} E^{a|J]} [E]$ is equivalent to the condition
\be \label{7}
S^{aIJ;bKL}:=\pi^{a[IJ}\;\; \pi^{|b|KL]}=0
\ee
provided that for any non zero vector $n$ the object ${Q^{ab}=
\pi^{aIK} \pi^{bJL} \delta_{IJ} n_K n_L}$ is non-degenerate. 
The proof follows closely the seminal paper \cite{FKP} in which 
the possibility of a higher dimensional, canonical version of LQG 
is also contemplated. The non-degeneracy of $Q^{ab}$ is equivalent to the non-degeneracy of $q_{ab}$, which is a standard restriction in canonical General Relativity, and we will keep this restriction in the classical theory in order to ensure equivalence with General Relativity. 

We are now in the position to establish the relation with the ADM phase space.
We postulate the following non-trivial Poisson brackets
\be \label{8}
\{A_{aIJ}(x),\pi^{bKL}(y)\}:=2 \beta \;G\; \delta_a^b \;
\delta_{[I}^K\;\delta_{J]}^L\; \delta(x,y) \text{,}
\ee
where $\beta$ is a free real parameter and 
consider the following quantities
\ba
\det(q)\;q^{ab}&:=&-{\rm Tr}(\pi^a \pi^b), \nonumber \\ 
\sqrt{\det(q)}\; K_a\;^b&:=&-\frac{1}{\beta} {\rm Tr}([A_a-\Gamma_a]\pi^b) \label{9}
\ea
which play the role of the intrinsic $D$ metric and the extrinsic curvature.
Then one can show \cite{I} that with the usual formula 
$P^{ab}:=\sqrt{\det(q)}q^{ac}[K_c\;^b-\delta_c^b \; K_d\;^d]$ one obtains
the following non-trivial Poisson brackets
\be \label{10} 
\{P^{ab}(x),q_{cd}(y)\}=-G\;\delta^{(a}_c\; \delta^{b)}_d\; \delta(x,y)
\ee
modulo terms that vanish on the joint constraint surface defined by the 
simplicity constraint (\ref{7}) and the so$(D+1)$ Gau{\ss} constraint
\be \label{11}
G^{IJ}=\partial_a \pi^{aIJ}+[A_a,\pi]^{IJ} \text{.}
\ee
This is non-trivial and heavily relies on the fact that the hybrid 
connection $\Gamma_{aIJ}[\pi]$ has a {\it weak potential}, that is, 
there is a functional $F[\pi]$ such that 
${\Gamma_{aIJ}(x)=\delta F/\delta \pi^{aIJ}(x)}$ modulo terms that vanish
on the simplicity constraint surface. Without this property, the 
bracket $\{P^{ab}(x),P^{cd}(y)\}$ would not vanish on the constraint surface.
It is also not difficult to show that the algebra of the simplicity and
Gau{\ss} constraints is first class and that the variables (\ref{9}) are 
weak Dirac observables with respect to both constraints. 

Thus, as with the usual LQG variables \cite{Ashtekar}, the (weak) 
integrability of the spin (hybrid) connection is central to show that 
the symplectic reduction of the SO$(D+1)$ Yang--Mills phase space by Gau{\ss} and 
simplicity constraints results in the ADM phase space. 
That the ADM spatial diffeomorphism constraints $C_a$ 
and Hamiltonian constraint $C$, when expressed in terms of (\ref{9}),
weakly Poisson commute with $S^{aIJ;bKL},\; G^{IJ}$ and that their algebra 
among themselves, as compared to the ADM phase space,
is unchanged modulo terms vanishing when 
$S^{aIJ;bKL}=G^{IJ}=0$ is a simple corollary. We conclude that we have 
found a connection formulation for Lorentzian General Relativity in $D+1$
dimensions with $D\ge 2$ with all the desired properties, in particular,
all four types of constraints form a first class algebra.

Remarkably, all we have said so far can be performed for all four
combinations of the space-time and internal signatures $(s,\zeta)$ respectively
where $s=\pm 1$ for Euclidean and Lorentzian General Relativity respectively
and ${\zeta=\pm 1}$ for SO$(D+1)$ and SO$(1,D)$ respectively.

Similar to the situation with the usual variables \cite{Ashtekar}, 
the constraints take a simple form when written in terms of the 
curvature $F_{abIJ}$ of $A_{aIJ}$. One finds modulo Gau{\ss} and 
simplicity constraints \cite{I} 
\ba \label{12}
C_a &=& -{\rm Tr}(F_{ab} \pi^b) \text{,} \nonumber
\\
\sqrt{\det(q)}C &=& -\zeta {\rm Tr}(F_{ab} \pi^a \pi^b) \nonumber \\
& & +\frac{1}{(D-1)^2}\; [\zeta-\frac{s}{\beta^2}]\;[K_a\;^b K_b\;^a-(K_c\;^c)^2] \nonumber \\
& &-K^{TT}_{aIJ}\; T^{aIJ;bKL}\; K^{TT}_{bKL} \text{.}
\ea
The spatial diffeomorphism constraint takes unsurprisingly a form analogous 
to the formulation in $D=3$. Also the first two terms in $C$ 
look familiar. However, the third term in the expression for $C$ is new.
Here $K_{aIJ}=(A_{aIJ}-\Gamma_{aIJ})/\beta$ and $K^{TT}_{aIJ}$ 
is transversal $n^I K^{TT}_{aIJ}=0$ and trace free $E^{aI} K^{TT}_{aIJ}=0$
on the solution $\pi^{aIJ}=2 n^{[I} E^{a|J]}$ of the simplicity constraint,
it can be written as $P_{aIJ}\;^{bKL}[\pi]\; K_{bKL}$ where $P[\pi]$ 
is a transverse tracefree projector and just depends on $\pi$. Also 
$T^{aIJ;bKL}[\pi]$ is a tensor constructed entirely from $\pi$. 
The significance of this third term is that it removes the dependence 
of the curvature on the transverse
tracefree components of $K_{aIJ}$ which are pure gauge with respect to 
the simplicity constraints and on which the ADM variables do not depend. 
Notice that for matching internal and 
space-time signature $\zeta=s$ and $\beta=1$ the second term in $C$
vanishes and the Hamiltonian constraint simplifies, a feature that also
is familiar from the usual formulation. In this case one can arrive 
at (\ref{12}) also starting from the Palatini action 
by the method of gauge unfixing \cite{II} where now the 
third term is generated when making the Hamiltonian constraint invariant
under the gauge transformations generated by the simplicity constraint
so that the afore mentioned second class partner can be dropped. 

In order to quantise the Hamiltonian constraint, we observe that the terms 
quadratic in $K_{aIJ}$ can be written, as in the $D=3$ situation \cite{QSD1} 
as \cite{III} 
\be \label{13}
K_{aIJ}(x)\propto \{A_{aIJ}(x),\{V,C_E\}\} \text{,}
\ee
where $V=\int\; d^Dx \sqrt{\det(q)}$ is the total volume and 
\be \label{14}
\sqrt{\det(q)}\; C_E:=-{\rm Tr}(F_{ab} \pi^a \pi^b)
\ee
is the ``Euclidean piece'' of the Hamiltonian constraint. Similar remarks 
apply to the tensors $T,P$ which can be treated by analogous Poisson bracket
identities as displayed in \cite{QSD1,QSD5}. Notice that the kinematical
Hilbert space techniques \cite{AI,AL,LOST,Fleischhack} as well as the 
treatment of the spatial diffeomorphism constraint \cite{ALMMT}
have been formulated 
for canonical theories of connections for compact gauge groups in any 
dimension and thus can be applied to our situation without further effort.
In particular, the holonomy--flux algebra now consists of SO$(D+1)$-valued 
holonomies of $A$ along piecewise analytic (or semianalytic) 1-dimensional 
 paths and so$(D+1)$-valued fluxes of $\pi$ through semianalytic $D-1$ surfaces. 

The following remark is due at this point for the case of $D=3$:\\
In the usual formulation the connection $A^{{\rm LQG}}$ is an SU(2) connection
and is related to the extrinsic curvature by 
\ba \label{15}
A^{{\rm LQG}}_{ajk}-\Gamma^{{\rm SPIN}}_{ajk}[E]
&=&\gamma \epsilon_{jkl} K^l_a; \nonumber \\
A^{{\rm LQG}}_{a0j}=\Gamma^{{\rm SPIN}}_{a0j}[E]&\equiv& 0 \text{,}
\ea
where $K_{ab}=K_a^j e_{bj}$ and where $\Gamma^{{\rm SPIN}}_{ajk}[E]$ 
is the spin connection 
of the co-triad $e_a^j$ built from $E^a_j$. Neither $A^{LQG}$ nor 
$\Gamma^{{\rm SPIN}}$ have a ``boost'' 
part, the information about the extrinsic curvature is encoded in the 
rotational part of $A^{{\rm LQG}}$ and depends on the Immirzi parameter 
$\gamma$. On the other hand, in the ``time gauge'' $n^I=\delta^I_0$ 
\ba
A^{{\rm NEW}}_{ajk}-\Gamma^{{\rm HYB}}_{ajk}[\pi]&\approx& S - {\rm gauge}, \nonumber \\
A^{{\rm NEW}}_{a0j}-\Gamma^{{\rm HYB}}_{a0j}[\pi]&=&\beta K_{aj} \text{,}  \label{16}
\ea
where $\Gamma^{{\rm HYB}}[\pi]$ is the hybrid connection built from
$\pi$. The tracefree rotational components of the new connection are 
pure gauge and the trace piece of the rotational components vanishes 
by the boost part of the Gau{\ss} constraint. Therefore the 
information about the extrinsic curvature is encoded in the boost part
of the new connection and depends on the parameter $\beta$.
It transpires that the Immirzi parameter $\gamma$ and $\beta$ have nothing 
to do with each other and that the new formulation is a rather different
extension of the ADM phase space as compared to the LQG formulation even
in $D=3$ which nevertheless have the same symplectic reduction, namely the 
ADM phase space. In \cite{II} we show that for $D=3$ one can generalise the 
exposition given so far by considering a 2-parameter family of connection
formulations depending on both $\gamma,\beta$. The essential features 
remain the same, the connection remains Poisson commuting. 

The price to pay for the higher dimensional connection 
formulation are the additional 
simplicity constraints. On the other hand, this makes this canonical 
formulation appear much closer to the spin foam models \cite{Perez}
which aim to be a path integral formulation of LQG in $D+1=4$ dimensions.
For instance, we expect that 
the $\beta=1,\zeta=s$ theory with arbitrary $\gamma$ corresponds to the
``new spin foam models'' \cite{EPRL,FK} because both can be obtained from 
the Holst modification of the Palatini action without imposing the time gauge.
Notice that our formulation is in the continuum rather than on a fixed 
triangulation. In particular, we must define the simplicity constraint
operators on spin network functions for all graphs not only simplicial ones.
The necessity of doing that in order to make contact with the canonical 
formulation has been recently emphasised also in \cite{KKL}. 

The
quadratic, canonical quantum simplicity constraints are naturally 
quantised by smearing the two factors of $\pi$ in (\ref{7}) over two 
independent $D-1$ surfaces in order to obtain flux operators which are then
shrunk to a point $x$ \cite{III}. Then two types of simplicity 
constraints arise: Either $x$ is an interior point of an edge (edge 
simplicity constraints) or a vertex (vertex simplicity constraints).
Taking advantage of the flexibility in the choice of the surfaces available
in the continuum formulation as well as taking linear combinations 
one can show that one can reduce the simplicity constraints to the 
following building blocks
\be \label{17}
S^{IJKL}_{e,e'}:=X^{[IJ}_e X^{KL]}_{e'}=0
\ee
for all pairs of edges $e,e'$ of the given graph which share a vertex.
Here $X^{IJ}_e$ is right invariant vector field on the copy of 
SO$(D+1)$ associated with the edge $e$. 
The edge simplicity constraints arise for $e=e'$, the vertex simplicity
constraints for $e\not=e'$. 
These are the analogs of the diagonal and cross simplicity constraints
appearing in the spin foam literature which involve either one face or 
two faces sharing a ``temporal edge''. These faces arise from our edges by 
``time evolving'' them into faces. 

The edge simplicity constraints 
enforce the irreducible SO$(D+1)$ representations on the edges of the 
\mbox{SO$(D+1)$} spin network functions to be simple representations which have 
already been described in great detail in \cite{FKP} for SO$(D+1)$. 
Despite the first class nature of the simplicity constraint in the classical, non-distributional theory\footnote{I.e. before using the singular smearing to construct holonomies and fluxed from the gauge and frame fields. In fact, the second class nature of the vertex simplicity constraints is already present in the classical theory when using a singular smearing, i.e. holonomies and fluxes. The wording ``anomaly'' however still seems to be appropriate, because the singular smearing is needed for the quantum theory, but not for the classical one.}, the vertex simplicity 
constraints become second class in the quantum theory due to the non-commutativity of the fluxes, which results in an anomaly and strong imposition leads to the higher 
dimensional analogue of the Barrett-Crane intertwiner also described 
already in \cite{FKP}. To avoid the anomaly one can invent several 
strategies, none of which are entirely satisfactory. Either one 
can try to construct a vertex master constraint \cite{III} for 
each vertex whose solutions, however, are beyond any analytic control
at the moment. 
Another possibility is to consider for each vertex a preferred recoupling 
scheme which then selects a maximal commuting subset of vertex 
simplicity constraints which in turn select a so-called simple intertwiner
\cite{V} which appears to establish a unitary map between the 
old and new formulation in $D+1=4$. 
The unnatural feature here is that the notion of simple intertwiner
depends on the chosen recoupling scheme. Yet another option in the 
spirit of the ``improved spin foam models'' \cite{EPRL,FK} is to consider
linear simplicity constraints \cite{V} instead. These arise in the 
canonical framework by considering the time normal $n^I$ as an independent
quantum field subject to the normalisation constraint ${\cal N}=n^I n_I-1$
and to replace the quadratic constraints (\ref{7}) by the linear
constraints 
\be \label{18}
S^{aIJK}=\pi^{a[IJ} n^{K]} \text{.}
\ee
The framework described so far can be adapted to this formulation of the 
simplicity constraint and all essential features are preserved. This formulation
automatically avoids the topological sector for $D=3$. The 
then needed Hilbert space representation for the time normal field
is supplied in \cite{VI}. Again, it is natural to smear the $\pi$ appearing
in (\ref{18}) over arbitrary surfaces. Notice that now the wave functions 
depend on both $A$ and $n$ where $A$ is located along the edges and $n$ along
the vertices. The building blocks of the simplicity constraint
are now the operators
\be \label{19}
X^{[IJ}_{e,x} n^{K]}(x) 
\ee
for each point $x$ on the edge $e$ and each edge $e$ of the graph. Here
$X_{e,x}$ denotes the right invariant vector field on the copy of SO$(D+1)$ 
corresponding to segment of $e$ which starts at $x$ and ends at the end 
point of $e$. The nice thing is that these constraints are non-anomalous
because they generate the Lie algebra of the SO$(D)$ subgroup
of SO$(D+1)$ stabilising $n$, however, they enforce that at each point of 
an edge there should be a specific $n$ dependence which maps out of the 
space of spin network functions described above. While some strategies for 
matching linear and quadratic simplicity constraints are spelled out 
in \cite{V} based on ideas from group averaging, none of them can be 
considered as complete so far and they deserve further research.

Notice that especially in $D=3$ we are now faced with a real challenge 
and opportunity to make progress with the simplicity constraint:\\
We have two classically equivalent connection formulations, the one 
based on the Ashtekar-Barbero-Immirzi SU(2) connection with SU(2)
Gau{\ss} constraint and the new SO(4) connection formulation with 
SO(4) Gau{\ss} and simplicity constraint. One would therefore expect the 
corresponding quantum theories to be unitarily or at least semiclassically
equivalent. Thus, the usual LQG formulation may help us to find 
a consistent and non-anomalous definition of the simplicity constraint
as argued in \cite{V}.

So far we have only considered the purely gravitational degrees of 
freedom. We now turn to matter. As far as standard matter 
is concerned, all the machinery developed for $D=3$ straightforwardly 
generalises to arbitrary $D$ as far as gauge bosons and scalars 
are concerned \cite{QSD5}. Slightly more attention is required when 
considering fermionic matter because the Lorentzian theory uses Dirac
matrices for the SO$(1,D)$ Clifford algebra which we would like to 
recast in terms of the Dirac matrices for the SO$(D+1)$ Clifford algebra 
in order that matter and gravitational degrees of freedom transform 
under the same gauge group SO$(D+1)$. As far as Dirac fermions are concerned,
one can proceed similarly \cite{IV} as in the case $D=3$ in \cite{QSD5}:
Consider the Lorentzian theory first in time gauge $n^I=\delta^I_0$ 
which breaks the SO$(1,D)$ internal group down to the SO$(D)$ subgroup
preserving the time gauge. In particular, the theory is now 
formulated in terms of $E^a_j,\; K_a^j$ as in \cite{Han}. 
Now extend all constraints and variables in an SO$(D+1)$ covariant way
such that they reduce to the previous expressions in the time gauge.
This can be done using that $E^{ai}=\pi^{ai0}$ in time gauge when the 
simplicity constraint holds, by subtracting terms proportional to 
$K^{TT}_{aIJ}$ as we did for the gravitational degrees of freedom and by 
replacing the Lorentzian $\gamma^0_L$ Dirac matrix by 
the Euclidean one $\gamma_0:=\gamma^0_E:=i\gamma^0_L$ wherever it appears.
For instance, the term $E^a_j \Psi^\dagger \gamma^0 \gamma_j \nabla_a\Psi$
where $\Psi$ are the Dirac fermions and $\nabla_a$ is the covariant derivative
defined by the SO$(D)$
spin connection of $E^a_i$, can be written as 
$\pi^{aIJ} \Psi^\dagger \gamma_{[I} \gamma_{J]} {\cal D}_a\Psi$
plus terms involving $K_{aIJ}$ where ${\cal D}_a$ is the covariant derivative 
defined by $A_{aIJ}$. Here we used the fact that the SO$(D)$ spin connection is 
the SO$(D+1)$ hybrid connection in time gauge. 
The $K_{aIJ}$ terms then need to be written as above
in double Poisson brackets involving volume and Euclidean piece 
of the gravitational contribution to the Hamiltonian constraint.
This produces 4-fermion terms, see \cite{IV} for details but apart from 
this one can copy the Hilbert space representation from the case $D=3$
for Dirac fermions given in \cite{KinematicalHS}. We conclude that 
standard matter can be treated just as in $D=3$.

The reason why this worked so well is that both SO$(D+1)$ and SO$(1,D)$ 
act on the same {\it complex} representation spaces. When we turn to 
Supergravity theories, this is no longer helps because Supergravity
multiplets contain the Rarita--Schwinger field (gravitino) as 
a {\it Majorana fermion} and these belong to {\it real} representation spaces.
In particular, the Lorentzian Dirac matrices (for our mostly plus signature 
choice in $D+1=4,10,11$) are real-valued which is crucial in order that the real vector space
$V$ formed by Majorana fermions is preserved under SO$(1,D)$. Since 
$\gamma^0_E=i\gamma^0_L$ is then purely imaginary this is no longer the case
for SO$(D+1)$ so that SO$(D+1)$ does not act on the vector space of 
Majorana fermions.
To see this explicitly, notice that $[\gamma^I,\gamma^J]/8$ are the 
generators of so$(1,D)$ or so$(D+1)$ respectively in the spinor representation.
We resolve the arising tension by noticing that while there 
is no action of SO$(D+1)$ on $V$ there is an action on the set of pairs
$(n,\theta)$ where $\theta$ is a Majorana fermion and $n$ is  the time normal 
field. To see this, notice that the action of $g\in$ SO$(D+1)$ on $n$ is simply given
by $n\mapsto g\cdot n$. Let us now build an SO$(D+1)$ matrix $A(n)$ from the 
components of $n$, for instance the one that is given in \cite{VI}, which
maps $n^I$ to the time gauge normal vector $\delta^I_0$.   
Then one can show that $g A(n)^{-1}=A(g\cdot n)^{-1} R(g,n)$ where $R(g,n)$ is a 
rotation preserving the time gauge. Such a rotation is generated in 
the spinor representation by $[\gamma^j,\gamma^k]/8,\;\;j,k=1,..,D$ 
and thus only involves the {\it real} Dirac matrices $\gamma^j$. We
now consider the extension $\theta\mapsto A(n)^{-1} \cdot\theta$ off the 
time gauge and observe that the SO$(D+1)$ orbit of these vectors 
is preserved as $g\cdot A(n)^{-1} \cdot\theta=A(g\cdot n)^{-1} \cdot R(g,n)\cdot \theta$.
In other words, we have the action on pairs $g\cdot (n,\theta)=(g\cdot n,
R(g,n)\cdot \theta)$. Starting from time gauge,
one then rewrites all constraints in terms of these 
variables, replaces Lorentzian $\gamma^0$ by the Euclidean one and 
finally extends off the time gauge similar as for Dirac fermions.
There is still a remaining complication which arises from the fact that 
there is a reality condition on the Majorana spinors which effectively 
gives rise to a non-trivial Dirac antibracket and which prohibits to use 
the formalism of \cite{KinematicalHS}. We supply the missing details 
and provide a proper Hilbert space representation of the Dirac antibracket
\cite{VI}.

Supergravity theories not only contain the Rarita--Schwinger field 
but often additional tensor fields such as the 3-index photon field $A$
in 11d Supergravity. This field is self-interacting due to a Chern-Simons 
term in the action (the Hamiltonian is a polynomial of fourth order 
in $A$ and the momentum $\pi$ conjugate to it) and it is constrained by a 
twisted Gau{\ss} law 
constraint which is of the form $G=d\ast \pi-c/2 F\wedge F$. Here,
$F=dA$ is the curvature of $A$ and $c$ is related to the level of the 
Chern--Simons theory. The appearance of the $F\wedge F$ term implies 
that one cannot simply solve the constraint by considering the theory
as the 3-form equivalent of Maxwell theory in higher dimensions and for which
gauge invariant states would simply be gauge invariant 
``Spin network states'', i.e. functions of $A$ 
integrated over closed 3-surfaces. In particular the analog of the 
LQG representation is inappropriate to solve the constraint. We 
resolve the tension \cite{VII} by considering the Weyl algebra generated by the 
exponentials of the Dirac observables $F,\; P=\ast \pi+c A\wedge F$
with respect to $G$ which can be computed in closed form and show that
the resulting $^\ast-$algebra admits a state (positive linear functional) 
which is discontinuous in both $F,P$ and is thus of the Narnhofer-Thirring
type\footnote{This state is different from the Fock state and has the attractive 
feature of being ghost free while manifestly Poincar\'e invariant.} \cite{NarnhoferThirring}. The resulting background independent Hilbert
space representation then follows from the GNS construction \cite{Haag} and since the 
contribution of the 3-index Photon to the Hamiltonian and spatial 
diffeomorphism constraint can be written just in terms of the observables,
they can be quantised in terms of Weyl elements \cite{VII}.\\
\\
We hope that the tools provided in our work will turn out to be useful for a rigorous canonical and non-perturbative quantisation, for instance by the methods of LQG, of a wide variety of Supergravity theories, including $d=4, N=8$, $d=11,N=1$ and $d=10, N=1$. An extension to the $d=10, N=2a$ and $d=10, N=2b$ theories related to type $2a$ and $2b$ Superstring theory should also be possible, although we are not aware of explicit Hamiltonian formulations of these theories. 

In order to attack the issue of comparing LQG techniques to String Theory, it makes sense to start with symmetry reduced models such as cosmology and black holes. The goal however should not just be to compare certain numerical values, like the black hole entropy, to each other, but to understand possible connections between the derivations of these. We hope that new features of higher dimensional (Super)LQG, e.g. supersymmetric effects in cosmological models and the treatment of topologically interesting black holes in higher dimensions (black rings), might yield hints on how to proceed. Also, despite many technical and conceptual hurdles along the way, revisiting the AdS/CFT correspondence \cite{AdSCFT} and its integrability structure \cite{Integrability} in the light of the proposed techniques seems to be an interesting project for further research.

\begin{acknowledgments} 
N.B. and A.T. thank the German National Merit Foundation for financial support. 
T.T. thanks Dieter L\"ust and Hermann Nicolai for motivational discussions about 
higher dimensional Ashtekar variables and their application to Supergravity and 
String Theory.
\end{acknowledgments}

\bibliographystyle{apsrev}

\end{document}